\documentclass[useAMS,usenatbib]{mnras}
\usepackage[T1]{fontenc}
\usepackage{ae,aecompl}
\usepackage{pdflscape}
\usepackage{graphicx}
\usepackage{longtable}
\usepackage{lscape}
\usepackage{amssymb}
\usepackage{endnotes} 
\usepackage{footnote}
\usepackage{subfigure}
\usepackage{amsmath}
\usepackage{txfonts}
\usepackage{natbib}
\usepackage{url}
\usepackage{breakurl}
\usepackage{helvet}
\usepackage{textcomp}
\usepackage{multirow}
\def\aj{{AJ}}
\def\araa{{ARA\&A}}
\def\apj{{ApJ}}
\def\apjl{{ApJ}}
\def\apjs{{ApJS}}
\def\aap{{A\&A}}
\def\aapr{{A\&A~Rev.}}
\def\aaps{{A\&AS}}
\def\jcap{{J. Cosmology Astropart. Phys.}}
\def\jrasc{{JRASC}}
\def\mnras{{MNRAS}}
\def\nar{{New A Rev.}}
\def\prd{{Phys.~Rev.~D}}
\def\pasp{{PASP}}
\def\jaa{{JoAA}}
\def\aipc{{AIP Conf. Ser}}
\def\aspconf{{ASP Conf. Ser.}}
\title[Supernova studies with the ILMT]{The zenithal 4-m International Liquid Mirror Telescope: a unique facility for supernova studies}

\author[B. Kumar et al.]
{Brajesh Kumar,$^{1}$\thanks{E-mail: 
\href{mailto:brajesh.kumar@iiap.res.in}{brajesh.kumar@iiap.res.in} (BK);} 
Kanhaiya L. Pandey,$^{1}$\thanks{
\href{mailto:kanhaiya.pandey@iiap.res.in}{kanhaiya.pandey@iiap.res.in} (KLP);} 
S. B. Pandey,$^{2}$\thanks{
\href{mailto:shashi@aries.res.in}{shashi@aries.res.in} (SBP)}
P. Hickson,$^{3}$
E. F. Borra,$^{4}$
\newauthor G. C. Anupama$^{1}$ 
and J. Surdej$^{5}$
\\\\
$^{1}$Indian Institute of Astrophysics, II Block, Koramangala, Bangalore 560 034, India\\
$^{2}$Aryabhatta Research Institute of Observational Sciences, Manora Peak, Nainital 263 002, India\\
$^{3}$Department of Physics and Astronomy, University of British Columbia, 6224 Agricultural Road, 
      Vancouver, BC V6T 1Z1, Canada\\
$^{4}$Department of Physics, Universit\'{e} Laval, 2325, rue de l'Universit\'{e}, Qu\'{e}bec, G1V 0A6, Canada\\
$^{5}$Institut d'Astrophysique et de G\'{e}ophysique, Universit\'{e} de Li\`{e}ge, All\'{e}e du 6 Ao\^{u}t 19, 
      B\^{a}t B5C, 4000 Li\`{e}ge, Belgium
}

\date{Accepted ------------, Received ------------; in original form ------------}
\pubyear{2018}

\begin{document}
\label{firstpage}
\pagerange{\pageref{firstpage}--\pageref{lastpage}}
\maketitle

\begin{abstract}

The 4-m International Liquid Mirror Telescope (ILMT) will soon become operational at the newly 
developed Devasthal observatory near Nainital (Uttarakhand, India). Coupled with a 4k $\times$ 4k 
pixels CCD detector and TDI optical corrector, it will reach approximately 22.8, 22.3 and 21.4 
magnitude in the $g'$, $r'$ and $i'$ spectral bands, respectively in a single scan. The limiting 
magnitudes can be further improved by co-adding the consecutive night images in particular filters. 
The uniqueness to observe the same sky region by looking towards the zenith direction every night, 
makes the ILMT a unique instrument to detect new supernovae (SNe) by applying the image subtraction 
technique. High cadence ($\sim$\,24 hours) observations will help to construct dense sampling 
multi-band SNe light curves. We discuss the importance of the ILMT facility in the context of SNe 
studies. Considering the various plausible cosmological parameters and observational constraints, 
we perform detailed calculations of the expected SNe rate that can be detected with the ILMT 
in different spectral bands. 

\end{abstract}


\section{Introduction}

The new generation large area survey programs have contributed with their invaluable data to discover 
new supernovae (SNe). In order to detect such transients, two different strategies are typically applied. 
First is the pointed survey approach where a fixed catalogue of galaxies are independently observed with 
a cadence of a few days. In the second method, a specific area of the sky is surveyed very frequently 
and transients are identified using the image subtraction technique. Some important SNe search programs 
based on these methods are: the Lick Observatory Supernova Search \citep[LOSS;][]{2000AIPC..522..103L},
the Panoramic Survey Telescope \& Rapid Response System \citep[Pan-STARRS;][]{2002SPIE.4836..154K},
the Dark Energy Survey \citep[DES;][]{2005astro.ph.10346T},
the Canada-France-Hawaii Telescope\,--\,Legacy Survey \citep[CFHT\,--\,LS;][]{2006A&A...447...31A},
the Carnegie Supernova Project \citep[CSP;][]{2006PASP..118....2H}
the Equation of State: SupErNovae trace Cosmic Expansion \citep[ESSENCE;][]{2007ApJ...666..674M,2007ApJ...666..694W},
the Sloan Digital Sky Survey \citep[SDSS;][]{2008AJ....135..338F,2008AJ....135..348S},
the Catalina Real-Time Transient Survey \citep[CRTS;][]{2009ApJ...696..870D},
the Palomar Transient Factory \citep[PTF;][]{2009PASP..121.1395L}, 
the All-Sky Automated Survey for SuperNovae \citep[ASAS-SN;][]{2014ApJ...788...48S} and the upcoming 
Zwicky Transient Facility \citep[ZTF;][]{2014htu..conf...27B}. 

It is notable that despite their great contribution to supernova (SN) search, this kind of projects are 
observationally expensive, requiring many hours of valuable telescope time to complete, and are also
depth and cadence limited. 
Furthermore, the majority of them do not perform observations of the same strip of
sky on a regular basis every night (however, see ASAS-SN survey, \url{http://www.astronomy.ohio-state.edu/asassn/index.shtml}).
The unconventional Liquid Mirror Telescopes \citep[LMTs,][]{1985PASP...97..454B,1992ApJ...393..829B,
1993PASP..105..501H} may provide a unique way to overcome some of these issues in a certain fashion. 
For SNe studies, LMT observations are useful over the generic facilities in several 
aspects: 

\begin{itemize}

\item Unbiased imaging:
Most nearby SNe are discovered by repeated imaging of catalogued galaxies \citep[][]{2001ASPC..246..121F} 
which introduces a possible bias towards metal rich galaxies. 
Though ongoing ASAS-SN survey like programs have improved the situation and much is expected with the 
upcoming ZTF facility. Moreover, a LMT will image the same strip of sky passing over it without any selection 
bias.

\item Inexpensive technology: 
The cost of constructing a moderate aperture telescope (4-m diameter) is roughly 1/50 that of a conventional 
telescope of the same class \citep{Borra_1a,2003A&A...404...47B}.

\item Continuous data flow:
There will be no loss of precious observing time because a LMT will observe continuously during the nights 
except for bad weather or technical problems and produce a large amount of scientific data using the sky light.

\item Easy image pre-processing:
Unlike conventional imaging, here image pre-processing is comparatively easier. For example, the image 
reduction is performed by dividing each column by a one-dimensional flat field. That can be achieved 
directly from the scientific data.

\item Deeper imaging:
Since each night the same sky strip will be captured by the telescope, we can co-add the consecutive night 
data to produce deeper images.

\end{itemize}

\begin{figure*}
\centering
\includegraphics[scale=0.135]{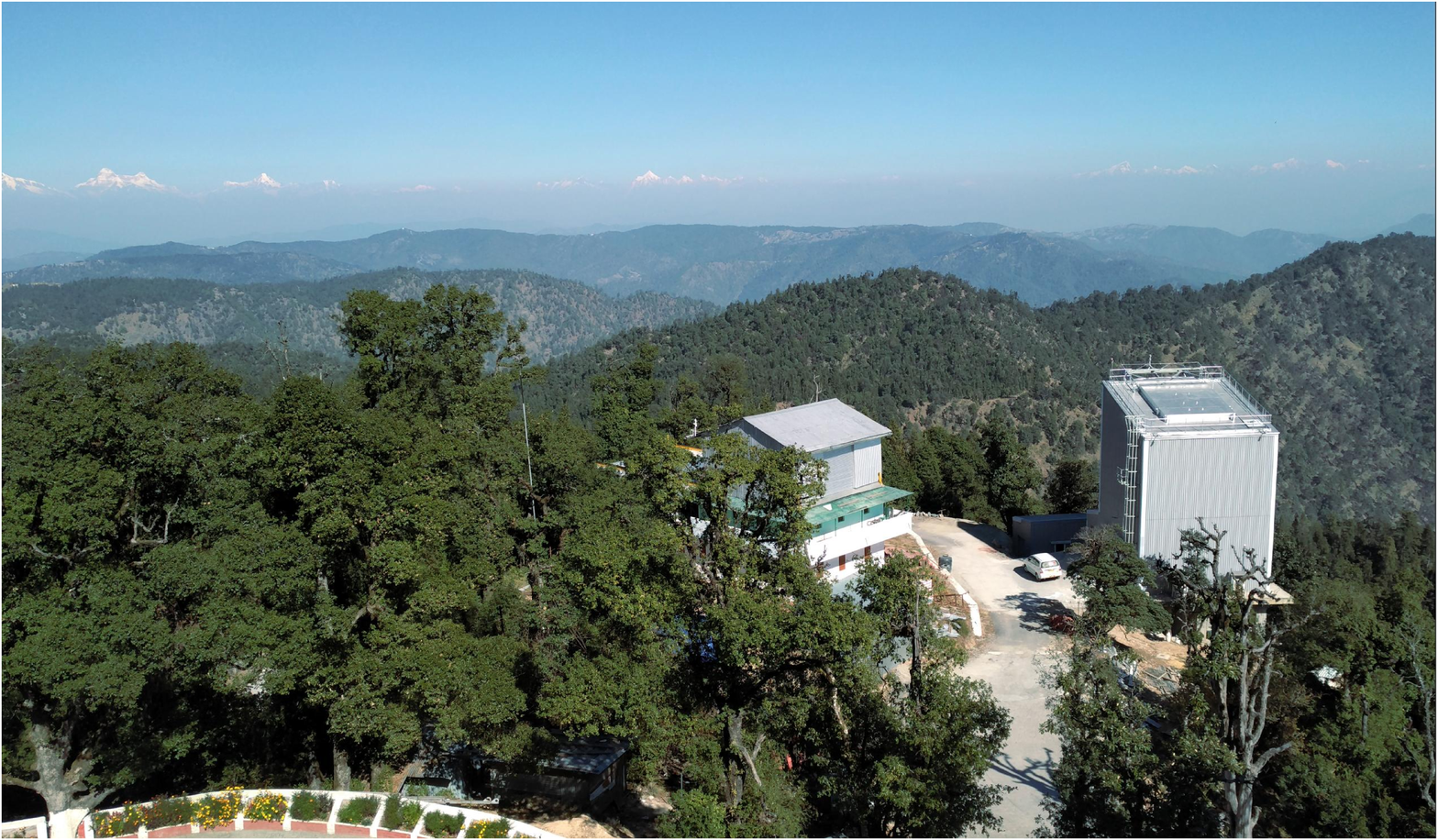}
\includegraphics[scale=0.11]{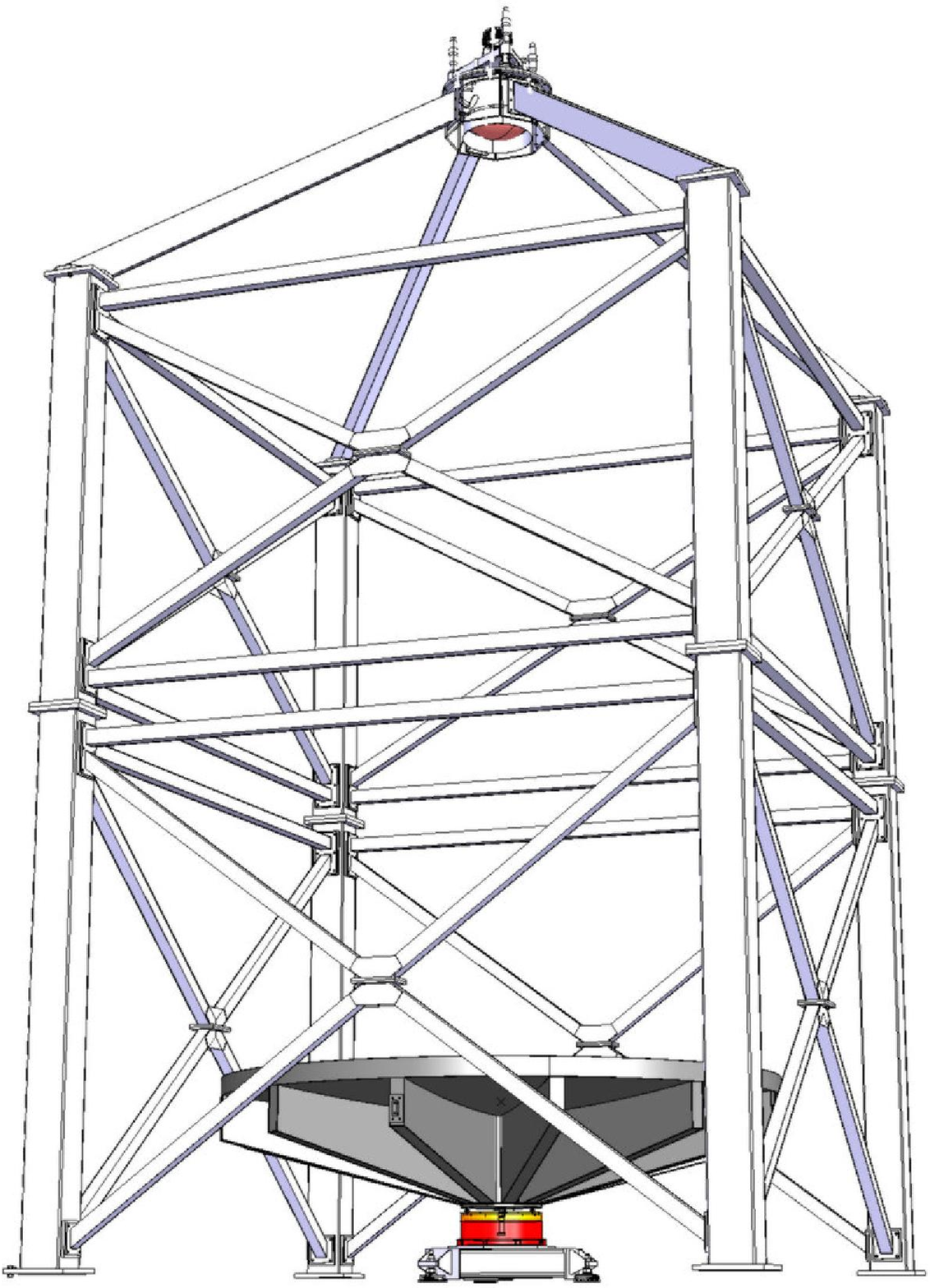}
\caption{Left panel: Panoramic view of the ILMT site. The 1.3-m \textit{DFOT} and the 4-m ILMT are
in the middle and right side, respectively in this image. Right panel: Major components of 
the ILMT. Here, the container is gray, the air bearing is red, the three-point mount (white) sits 
below the air bearing and the vertical steel frames (white) hold the corrector and the CCD camera at
the top. The tentative size and other parameters of the telescope are listed in Table~\ref{ilmt_lim}.
Note the nice view on the Himalayan chain in the background of the left photograph.}
\label{ILMT_pan}
\end{figure*}

\begin{figure}
\centering
\includegraphics[scale=0.27]{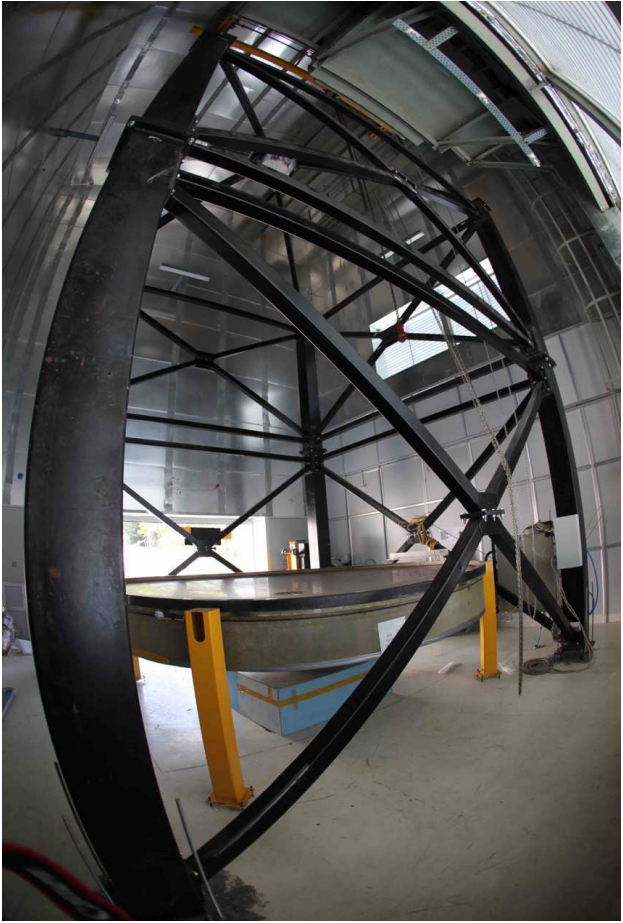}
\caption{Fish eye view of the present status of the ILMT.
To protect from the dust, the air bearing and the three-point mount are covered with a wooden box 
(blue colour). Four safety pillars (yellow colour) are also visible near the parabolic container 
to prevent any mercury spill.}
\label{ILMT_pan2}
\end{figure}

Different size LMTs have already been built and operated by several groups \citep[for example;][]
{1989ApJ...346L..41B,1994ApJ...436L.201H} and the 6-m diameter Large Zenithal Telescope 
\citep{2007PASP..119..444H} was the largest one in its class. The scientific contributions 
of these facilities were mainly limited due to the lack of an appropriate TDI corrector 
and/or large CCD camera and/or location (c.f. poor sky conditions). Therefore, a full time 
LMT entirely dedicated to astronomical observations was proposed and the idea of building 
the International Liquid Mirror Telescope (ILMT\footnote{\url{http://www.ilmt.ulg.ac.be}}) 
was born. SNe related study is one of the major scientific interests behind the ILMT project
\citep{2006SPIE.6267E...4S}.

A link between the star formation history and the cosmic supernova rate has been an open question.
It is generally believed that Type Ia SNe originate from the intermediate mass (3\,--\,8 M$_{\odot}$) 
and both young and old population stars \citep*[][for recent reviews]{2012NewAR..56..122W,
2014ARA&A..52..107M}. Such explosions are supposed to be the consequences of 
thermonuclear disruptions when a carbon-oxygen white dwarf reaches the critical Chandrasekhar mass 
limit \citep[$\simeq$1.4 M$_{\odot}$,][]{1931ApJ....74...81C} by accreting matter from its evolving 
binary companion \citep*{1960ApJ...132..565H,1973ApJ...186.1007W,1986ApJ...301..601W,2000ARA&A..38..191H}. 
The observational features of these events show homogeneity and due to their high luminosity near 
the maximum light, they are detectable at high redshift \citep{2005ASSL..332...97F}. Type Ia SNe are 
considered to be reliable standard candles \citep{1992ARA&A..30..359B,1993ApJ...413L.105P} and play 
an important role to constrain the geometry of the Universe \citep[e.g.][]{1998AJ....116.1009R,
1999ApJ...517..565P,2007ApJ...659...98R}.

Contrary to the progenitors of Type Ia SNe, massive stars ($M \ge\, $8 M$_{\odot}$) follow a different 
evolutionary path \citep[e.g.][]{1991ComAp..15..221B,2003ApJ...591..288H,Smartt2009,2012ARA&A..50..107L}.
At the final stage when their nuclear fuel is exhausted, the gravitational collapse of the stellar core 
triggers into a catastrophic death that appears in the form of a core-collapse supernova (CCSN).
The spectro-photometric features led to classify them in several types \citep[c.f. IIP, IIL, IIn, IIb, Ib, 
Ic and Ic-BL; see][for a review]{1941PASP...53..224M,1997ARA&A..35..309F}.
CCSNe show diverse observational properties.
For example, their absolute magnitude distribution peaks roughly 1.5 mag fainter than SN Ia and covers a 
range of more than 5 mag \citep{2002AJ....123..745R,2006AJ....131.2233R}. Similarly, a wide dispersion is 
seen in the ejecta mass, kinetic energy, radiated energy and the amount of synthesized radioactive materials 
created in the explosion. This indicates that possibly different physical mechanisms play an important role 
during the evolution phases of the progenitors such as stellar wind \citep*{2008A&ARv..16..209P}, mass 
transfer in a binary system \citep*{1985ApJ...294L..17W,1992ApJ...391..246P,2010ApJ...725..940Y,
2012Sci...337..444S} and mass loss \citep{2006ApJ...645L..45S,2014ARA&A..52..487S}, etc.

Along with astrophysical and cosmological implications, SNe are also primarily responsible for the 
chemical enrichment of galaxies through their heavy elements and dust \citep[e.g.][]{1986A&A...154..279M,
2001MNRAS.325..726T,2007MNRAS.378..973B}. Furthermore, the expanding shock waves produced during the 
explosion sweep, compress and heat the surrounding interstellar medium that finally triggers the star 
formation process \citep[e.g.][and references therein]{1977ApJ...217..473H,1998ASPC..148..150E}.
Unbiased and large sample studies of SNe may provide answers to some of the underlying questions 
related to the star formation history, progenitor evolution scenario and parameters causing the 
diversity in their observed properties. In this context the ILMT deep imaging survey along 
with complementary observations from other existing observational facilities will be advantageous.


\section{The ILMT project}\label{project}

The ILMT project is a scientific collaboration between four countries: Belgium, India, Canada 
and Poland. The main participating institutions are: the Li\`ege Institute of Astrophysics and 
Geophysics (University of Li\`ege, Belgium), the Aryabhatta Research Institute of Observational 
Sciences (ARIES, India), the Royal Observatory of Belgium, several Canadian universities (British 
Columbia, Laval, Montr\'eal, Toronto, Victoria and York) and the Observatory of Pozna\'n (Poland). 
The AMOS (Advanced Mechanical and Optical Systems) company in Belgium has participated in the 
manufacturing of the telescope.

The ILMT is being installed at the Devasthal (meaning `Abode of God') mountain peak, in the
central Himalayan range in India. This is a newly developed observatory under ARIES. A panaromic 
view of the site is illustrated in Fig.~\ref{ILMT_pan}. The Devasthal observatory is situated at 
an altitude of $\sim$2450m, with longitude $79^{\circ}$ $41^{\prime}$ $04^{\prime\prime}$ East 
and latitude $+29^{\circ}$ $21^{\prime}$ $40^{\prime\prime}$ \citep[c.f.][]{Sagar2011Csi...101...8.25,
2012ASInC...4..173S}.
It is important to highlight that in view of the site advantages, apart from the upcoming 
4-m ILMT, the other existing astronomical observing facilities at Devasthal are the 1.3-m 
{\it DFOT}\footnote{Devasthal Fast Optical Telescope} and 3.6-m {\it DOT}\footnote{Devasthal 
Optical Telescope}. Major scientific prospectives of these telescopes can be found in 
\citet[and references therein]{2016arXiv160706455S}.

A sketch of the ILMT structure is shown in Fig.~\ref{ILMT_pan} (right panel). It consists of three 
major parts, namely the air bearing, the container and the vertical structure which will hold the
corrector and CCD camera. The primary mirror is a 4-m diameter epoxy-carbon-fiber structure that 
has a smooth parabolic upper surface produced by spin casting \citep{Magette2010,Kumar2014}, 
see Fig.~\ref{ILMT_pan2}.
The dish will support a thin layer (approximately 2\,--\,3 mm thick) of liquid mercury that will produce 
the reflecting surface.  When the mirror is rotated uniformly about its vertical axis, the combination 
of gravity and centrifugal force will produce an equilibrium surface that is parabolic to high accuracy.
A detailed explanation can be found in \citet{1982JRASC..76..245B}.
Although mercury vapour is harmful, it is greatly suppressed by a thin transparent layer of oxide 
that forms soon after emplacement. Moreover, a thin film of mylar\footnote{It is a scientific grade 
polyester film (thickness $<$\,12 $\mu$m). Optical quality tests of such films for the LMTs are 
discussed in \citet{1992PASP..104.1239B,2007PASP..119..456H}.}, co-rotating with the mirror, will 
contain any remaining vapour. This film is required to prevent vortices, produced in the boundary 
layer above the mirror due to its rotation, from disturbing the liquid surface.

The ILMT is a zenithal rotating telescope. It cannot track stellar objects like conventional 
glass mirror telescopes. Therefore, images are secured by electronically stepping the relevant CCD charges. 
The transfer rate is kept similar as the target drifts across the detector (i.e. equal to the observatory 
sidereal rate). This specific technique is known as the Time Delayed Integration (TDI) or drift-scanning 
\citep[see][and references therein]{1992MNRAS.258..543G}. Advantages of the TDI mode of observations 
can be found in \citet[][and references therein]{Kumar2014}. Because the primary mirror is parabolic, a 
glass corrector will be used to obtain a good image quality over a field of view of 27 arcmin in diameter 
including TDI correction \citep[see][]{1998PASP..110.1081H,2002A&A...388..712V}.
A CCD detector ($4096\times4096$ pixels) manufactured by `Spectral Instruments' will be positioned at 
the prime focus, located about 8m above the mirror. The ILMT observations will be mainly performed with 
the $i'$ filter (although there are additional filters $g'$ and $r'$). This will be advantageous 
for a maximum number of nights because the spectral range covered by the $i'$ filter is less sensitive 
to the bright phases of the moon. Initially the ILMT project will be for 5 years which will allow us to
collect a large sample of stellar objects including transients.
More detailed information about its instruments and science cases can be found 
elsewhere \citep[e.g.][and references therein]{2006SPIE.6267E...4S,Jean_bina,Magette2010,
2012IAUS..285..394P,Finet2013,Kumar2014,2015ASInC..12..149K,Kumar_bina}.

During the last few years, several experiments have been performed to sort out technical difficulties
related to the ILMT. In continuation of such activities we have also performed TDI mode observations 
from the Devasthal observatory using the 1.3m DFOT. These images have been used to test the ILMT data 
reduction pipeline and preliminary results are presented in \citet{Pradhan_bina}. The installation 
process of the telescope began in 2017 March and is now in its final stage. 
The metallic structure (to hold the CCD camera and corrector), safety pillars, air bearing are already 
erected. To ensure optimal and very safe operation of the air bearing two parallel air supply systems 
have been installed. In addition, several components/instruments like pneumatic valves, air dryers, 
air filters and sensors (pressure, temperature, humidity and dew-point) are also installed. 
First light of the ILMT is expected before the beginning of the 2018 {\it Monsoon} season. 
For the present status of the ILMT project, see \citet{Jean_bina}.
A fish eye view of the installation is shown in Fig.~\ref{ILMT_pan2}. 


\section{ILMT limiting magnitudes and accessible sky area}\label{Estimation1}

The scientific performance of an instrument depends on the maximization of its throughput.
Considering various parameters (e.g. transmission coefficients from the mirrors, filters, 
CCD glass, sky, extinction and quantum efficiency of the CCD chip), the expected counts ($N_{\rm e}$) 
from a star of certain magnitude (m) can be estimated using the following formula 
\citep{1989ecaa.book.....M,1991JApA...12..319M}.   

\begin{equation}
N_{\rm e} = 3.95 \times 10^{11} ~ D^{2} ~ \lambda_{\rm n} ~ \Delta\lambda_{\rm n} ~ F_{0}^{\rm n} ~ 10^{-0.4{\rm m}} 
A_{\rm F} ~ \eta 
\end{equation}

where D is the diameter of the telescope, $\lambda_{\rm n}$ and $\Delta\lambda_{\rm n}$ are
the effective wavelength and bandwidth of the filters, $F_{0}^{\rm n}$ is the flux density 
(per wavelength) from a star of magnitude 0 at the wavelength $\lambda_{\rm n}$ above 
the Earth atmosphere, $A_{\rm F}$ is the fractional reflecting area of the mirror surface and 
$\eta$ is the efficiency of the system (mirror + filter + CCD).

Assuming that each optical photon is capable of producing a corresponding photo-electron, 
the full well capacity of the required CCD pixel could be estimated by assuming a certain 
integration time for a star with a known brightness. 
Furthermore, if the sky brightness is known for a given CCD, we can also calculate the sky 
counts and the underlying noise. 

\begin{equation}
N  = \sqrt{(N_{\rm e} ~ e_{\rm t} + S_{\rm e} ~ e_{\rm t} ~ n_{\rm p} + D_{\rm c} ~ e_{\rm t} ~ n_{\rm p} 
+ R_{\rm n}^2 ~ n_{\rm p})}
\end{equation}

Here, 
$N_{\rm e}$ indicates the number of electrons (per sec), $e_{\rm t}$ the exposure time (sec), 
$S_{\rm e}$ the sky brightness (in electrons), $n_{\rm p}$ the number of pixels in the image of the 
observed star, $D_{\rm c}$ the dark current (e$^-$/pix/sec) and $R_{\rm n}$ the read out noise.

The signal-to-noise ratio (SNR) can also be calculated for stars with different brightness 
\citep{1989ecaa.book.....M}. 

\begin{equation}
{\rm SNR} = \left( \frac {{N_e \times e_t}}{N}\right). 
\end{equation}

The CCD readout noise is Gaussian while the star counts, dark counts are Poisson in nature. 
The aperture to calculate the star light is considered as circular. For the present calculations, 
the FWHM is considered as 1.5 arcsec, nearly equal to the median seeing at Devasthal.
The optimal aperture is considered to be 1 $\times$ FWHM \citep[see][]{1989PASP..101..616H,Howell}.
We can also estimate the corresponding error in the magnitude estimation by knowing 
the value of the SNR \citep{2011A&A...531A.151D} 

\begin{equation}
\sigma_{mag} = 2.5 \times log_{10}[1+1/{\rm SNR}].
\end{equation}

\begin{table}
\centering
\small
\caption{Different parameters used to calculate the ILMT limiting magnitude.
See also \citet{Finet2013}. \label{ilmt_lim}}
\begin{tabular}{ll}
\hline 
Diameter                             & 4.0-m      \\
Fraction of reflecting area          & 0.95       \\
Reflectivity                         & 0.77       \\
Mylar transmission                   & 0.80       \\
Corrector transmission               & 0.85       \\
FWHM                                 & 1.5\arcsec       \\
CCD pixel size                       & 0.4\arcsec/pixel \\
CCD dark noise                       & 0.00083 e$^{-}$/pixel/sec \\
CCD readout noise                    & 5.0 e$^{-}$ \\
CCD gain                             & 4.0 e$^{-}$/ADU \\
Wavelength ($g', r', i'$)            & 4750, 6250, 7630 \AA \\
Wavelength FWHM ($g', r', i'$)       & 1450, 1500, 1500 \AA \\
Extinction ($\sim$$g'$,$\sim$$r'$,$\sim$$i'$) & 0.21, 0.13, 0.08 mag \\
Sky mag    ($\sim$$g'$,$\sim$$r'$,$\sim$$i'$) & 21.3, 20.5, 18.9 mag/arcsec$^{2}$\\
CCD quantum efficiency ($g', r', i'$)& 0.70, 0.91, 0.91  \\
Filter transmission   ($g', r', i'$) & 0.92, 0.95, 0.95  \\
System efficiency, $\eta$ ($g', r', i'$)  & 0.55, 0.74, 0.74  \\  
\hline
\end{tabular}
\end{table}

\begin{figure}
\centering
\includegraphics[width=\columnwidth]{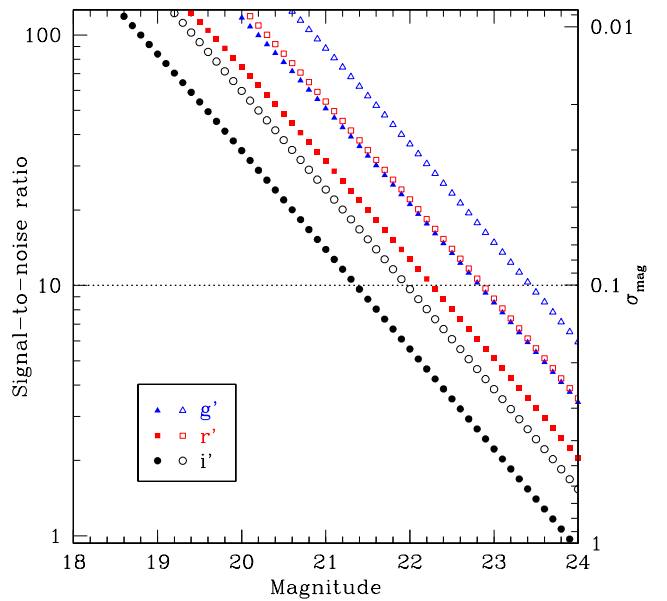}
\caption{
The ILMT limiting magnitudes for the $g'$, $r'$ and $i'$ filters are shown with different symbols.
The X-axis represents the limiting magnitude and the Y-axis represents the SNR and the corresponding
error in magnitude. The filled and open symbols represent the results for the exposure of a single
scan (102 sec) and three scans (306 sec), respectively (see Section~\ref{Estimation1} for details).
The dotted horizontal line is indicative for a SNR of 10 and an uncertainty of 0.1 mag.
Approximately 0.5 mag is gained once we stack images taken over three nights in any single filter.}
\label{ilmtlimit}
\end{figure}

We have estimated the limiting magnitudes of the ILMT for different filters ($g'$, $r'$ and $i'$) 
using the above equations. The various parameters used for these estimations are listed in 
Table~\ref{ilmt_lim}. The limiting magnitudes estimated using the above methods for different 
filters are plotted in Fig.~\ref{ilmtlimit} with different symbols. It is obvious from this figure 
that with an exposure time of 102 sec, the limiting magnitudes are $\sim$22.8, $\sim$22.3 and 
$\sim$21.4 mag for the $g'$, $r'$ and $i'$ filters, respectively. Furthermore, since during
each night the same strip of sky will pass over the telescope (except for a 4 min. shift in 
right ascension); successive night images can then be co-added. This will yield a 
longer effective integration time. Therefore, we have also estimated the limiting magnitudes 
for 306 sec (3 night images) exposure time, using the same parameters. The estimated magnitude 
limit improves to $\sim$23.4, $\sim$22.8 and $\sim$22.0 mag for the $g'$, $r'$ and $i'$ filters, 
respectively.
The co-addition technique is not limited only for 3 nights but it can also be applied for several
more night imaging data. Consequently, we may reach very faint magnitude levels \citep[see also][]{Borra_1a,
Borra_1b, 2003A&A...404...47B}. 


Pointing towards the zenith, the ILMT field-of-view (FOV) is centered at the Devasthal observatory 
latitude which is 29.36$^\circ$ N. The ILMT FOV is approximately 27\arcmin\ by 27\arcmin. 
One can find that the total accessible sky area with the ILMT will be 141.2 square degrees.
Out of it only 1/3 nightly strip ($\sim$47 square degrees) can be monitored.  
At high galactic latitudes ($\mid$b$\mid$ $>$ 30$^\circ$) the detection of fainter and more distant 
objects (e.g. SNe, galaxies, quasars ...) will be possible \citep[see][]{2006SPIE.6267E...4S,
Magette2010,Finet2013,Kumar2014}.
%
\begin{figure}
\centering
\includegraphics[scale=0.85]{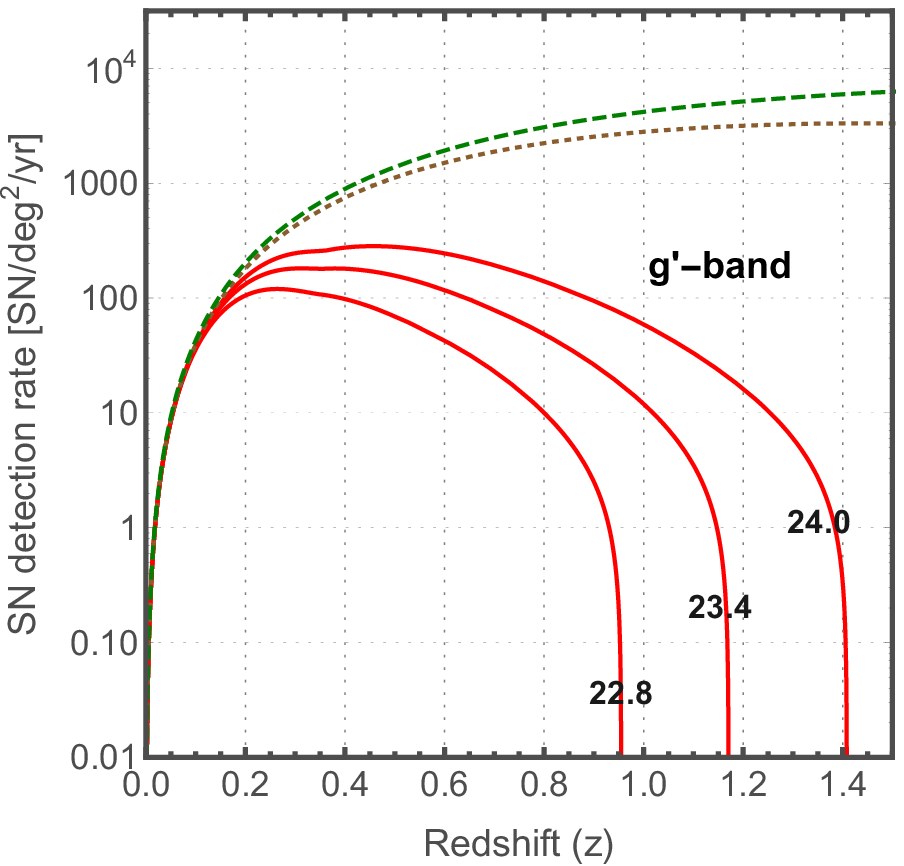}
\includegraphics[scale=0.85]{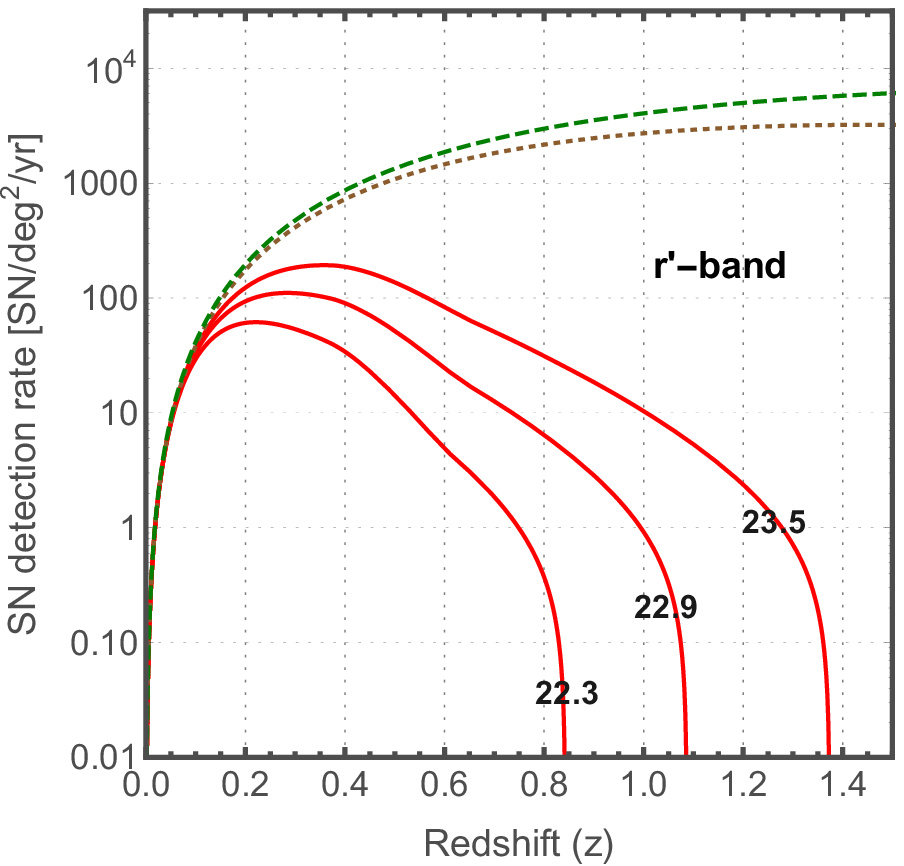}
\includegraphics[scale=0.85]{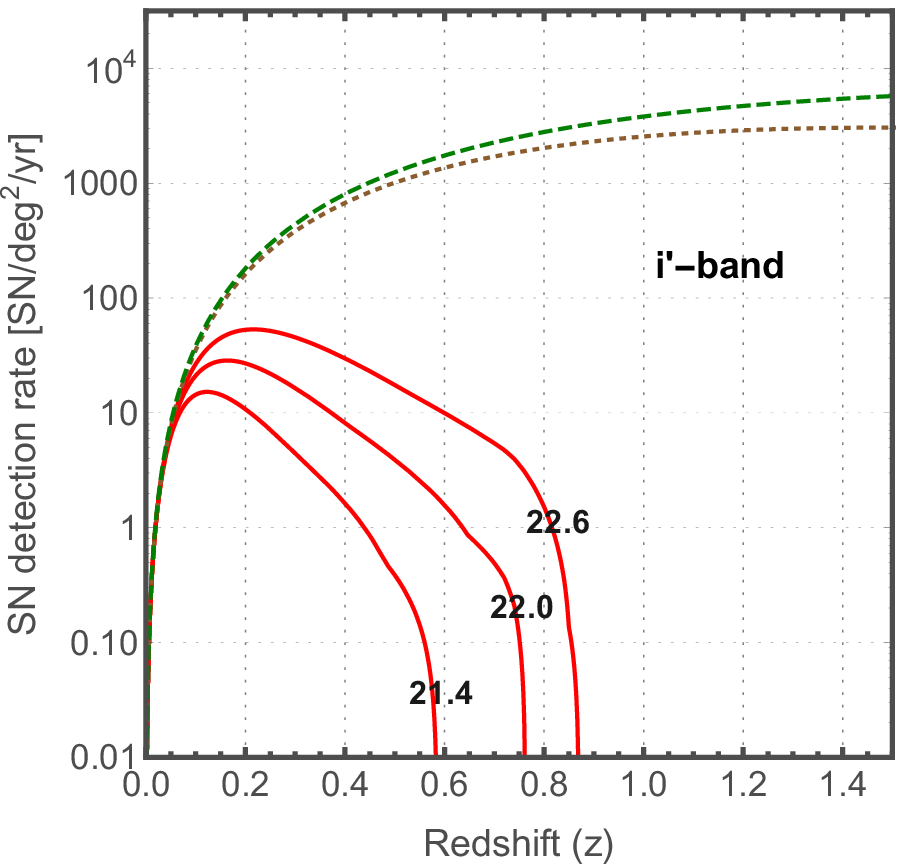}
\caption{Detection rate of CCSN as a function of redshift.
The dashed (green colour) and dotted (brown colour) curves, respectively indicate the cosmic CCSN rate
without and with dust extinction consideration. Possible number of CCSNe to be detected with the ILMT
in different bands ($g', r'$ and $i'$) and for different magnitude limits (c.f. stacking of
consecutive night images, see Section~\ref{Estimation1}) are also shown.}
\label{fig_snr_cc}
\end{figure}

\begin{figure}
\centering
\includegraphics[scale=0.85]{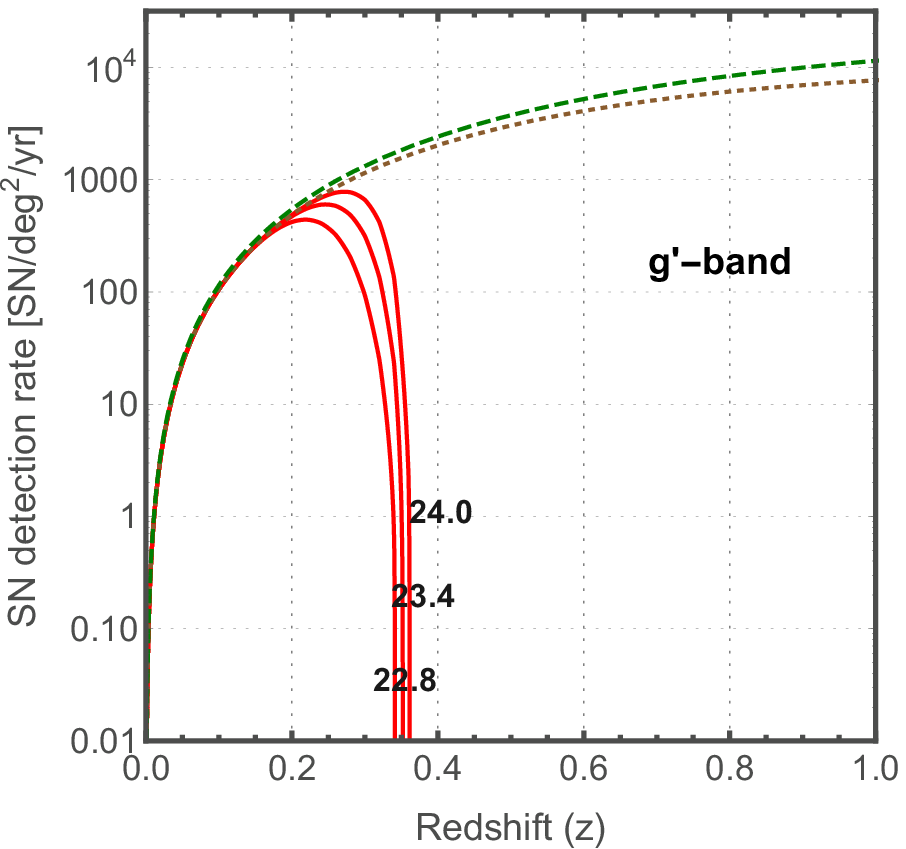}
\includegraphics[scale=0.85]{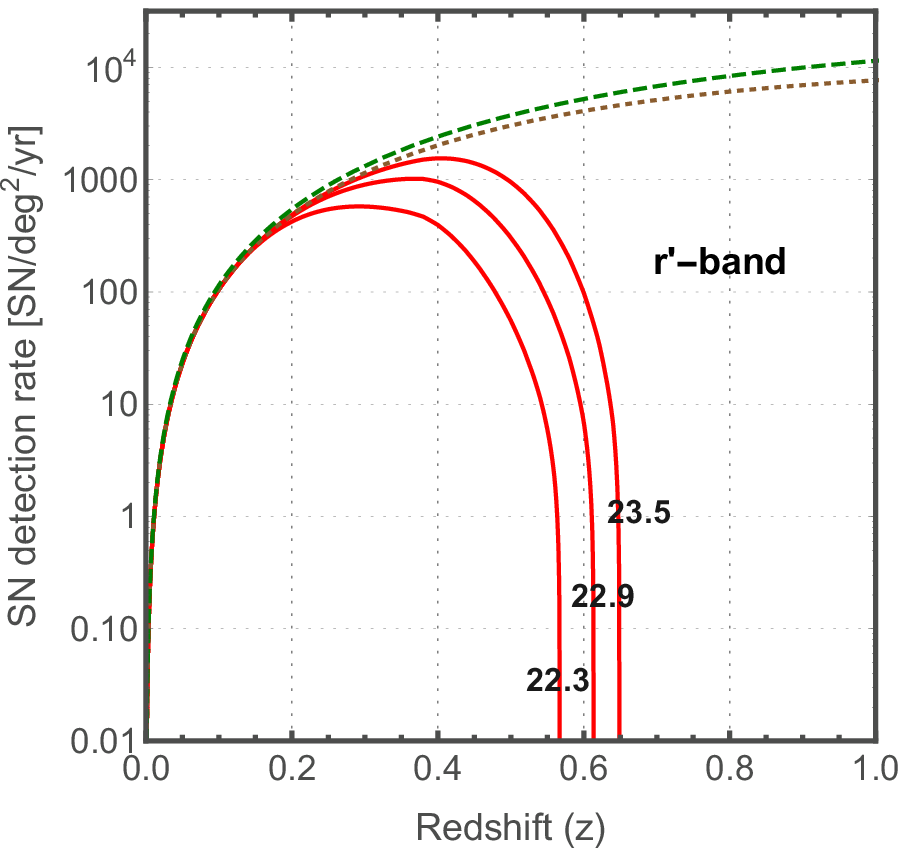}
\includegraphics[scale=0.85]{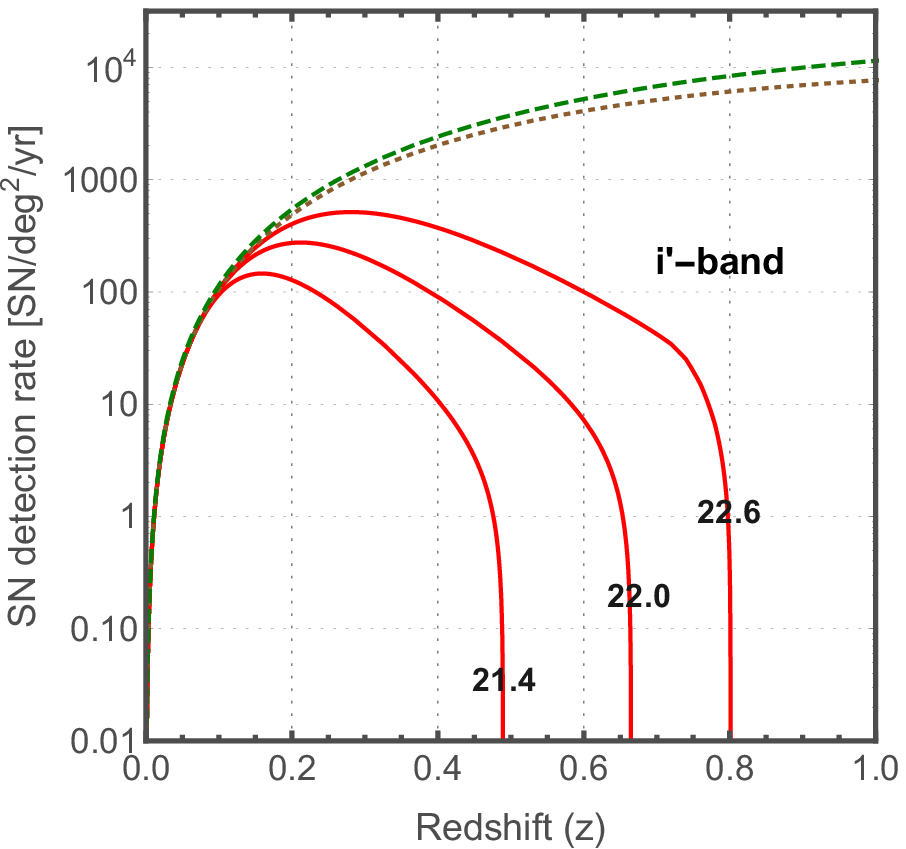}
\caption{Detection rate of Type Ia SN as a function of redshift. The curves are similar to
those in Fig.~\ref{fig_snr_cc} but for Type Ia SNe.}
\label{fig_snr_1a}
\end{figure}

\section{Supernova rate and ILMT}\label{sn_rate}

The astronomical community is deeply interested in understanding the nature of the different kinds of 
SNe and their evolution with redshift. The CCSNe rate is expected to reflect the star-formation rate, 
increasing with redshift as (1\,+\,$z$)$^{\beta}$ (for $z$$\,\approx$\,0.5) where $\beta$ is in the range
2.5 to 3.9 \citep[see][]{2004ApJ...615..209H,2005ApJ...632..169L, 2005ApJ...619L..47S,2006ApJ...651..142H,
2010ApJ...718.1171R,2012A&A...539A..31C}.
The type Ia SNe rate rise is rather slow with redshift, $\sim$(1 + $z$)$^{\beta}$ \citep[see][and references
therein]{2008ApJ...683L..25P,2012AJ....144...59P}, where $\beta$ is $2.11 \pm 0.28$ up to $z$ $\sim$1.
In order to search for a possible correlation between the star formation and SN rates in the local 
universe, several studies have been performed \citep[see, e.g.][and references therein]{2004ApJ...613..189D,
2006AJ....132.1126N,2008ApJ...682..262D,2011MNRAS.417..916G,2014ApJ...792..135T,2015A&A...584A..62C,
2017A&A...598A..50B}.

In the framework of supernova studies with liquid mirror telescopes, \citet{Borra_1a,Borra_1b,
2003A&A...404...47B} has described the cosmological implications of SNe and estimated the number of 
SNe for a strip of sky using the expected rate of SNe given in \citet{1996ApJ...473..356P}. In the 
following we have performed a detailed calculation of the expected number of SN events which can be 
detected with the ILMT. We calculate the detection rate for the core-collapse and Type Ia SNe for all 
the three proposed bands of the ILMT (see Figs.~\ref{fig_snr_cc} and \ref{fig_snr_1a}). For the 
calculations we follow the prescription given in \citet{2009JCAP...01..047L}. In the following we 
present a brief description of the steps and the quantities involved in the calculations.

The supernova detection rate per unit redshift per unit solid angle in a filter band $x$ 
can be expressed as follows,
\begin{equation}
\frac{dN_{SN,obs,x}}{dt_{obs}dzd\Omega} = R_{\rm SN}(z) \ f_{detect}(z;m_{lim, x}^{\rm SN}) \
\frac{r(z)^2}{1+z} \frac{dr}{dz} 
\end{equation}
where $r(z)$ is the co-moving distance, $R_{SN}(z)$ is the cosmic SN rate which can be 
written as 
\begin{equation}
R_{\rm SN}(z) = \frac{X_{\rm SN}}{\left<m\right>_{\rm SN}} \ \dot\rho_\star(z)
\end{equation}
where $X_{SN}$ is the fraction of stellar mass which results in supernovae
(=$\int_{SN} M \xi(M) dM \bigg/ \int M \ \xi(M) dM)$, $\left<m\right>_{\rm SN}$ is the average supernova 
progenitor mass  (=$\int_{SN} M \ \xi(M) dM \bigg/ \int \xi(M) dM)$ and $\dot\rho_\star(z)$ is the star 
formation rate. $\xi(M)$ represents the stellar initial mass function (IMF).
The quantity $f_{detect}(z)$ is the fraction of SNe which can be detected by the instrument and 
it depends on the characteristics of the instrument and the SN type ($i$), 
\begin{equation}
f_{detect}(z) =  f_{dust}(z) \times \frac{1}{N} \sum_{i=1}^{N} \int^{m_{lim,x}}_{-\infty} \phi_i(m, z) \ dm 
\end{equation}
$i$ denotes different SN types, for example in case of CCSNe we considered all four types, 
Ibc (both Ib and Ic), IIL, IIP and IIn (i.e. N = 4), whereas for Type Ia SNe, N = 1. 
The $\phi_i(m,z)$ is the supernova luminosity function or the magnitude distribution function which is assumed 
to be a normal distribution with the mean magnitude $\tilde{m}_i$ and variance $\sigma_i$, 
where $\tilde{m}_i = \tilde{m}_i^{abs} + 5 \log[{d_L(z)}/10\, \text{pc}] + K_{ix}(z) + \eta_{ixB} 
+ A(z)$; $d_L(z) $ is the luminosity distance, $K_{ix}(z)$ is the $K$-correction, $\eta_{ixB}$ is the color 
correction and $A(z)$ is the dust correction \citep[for details see,][]{2009JCAP...01..047L}. 

The $K$-correction and the color correction can be computed as follows,
\begin{eqnarray}
\begin{aligned}
K_{ix}(z) &=& 2.5\log(1+z) \ + \ 2.5\log \frac{\int F_i(\lambda) \ S_x(\lambda) d\lambda}{\int F_i(\lambda/(1+z)) \ S_x(\lambda) d\lambda} 
\end{aligned} \\
\begin{aligned}
\eta_{ixB} &=& -2.5\log \frac{\int F_i(\lambda) \ S_x(\lambda) d\lambda}{\int F_i(\lambda) \ S_B(\lambda) d\lambda} + {\rm zeropoint \ correction}
\end{aligned}
\end{eqnarray}
here, $F(\lambda)$ is the intrinsic spectral distribution of the supernova, $S_x(\lambda)$ and $S_B(\lambda)$ are 
the response functions of the filter $x$ (ILMT filter bands) and $B$ (filter band used to estimate the absolute 
magnitude distribution), respectively \citep[for more details see][]{2009JCAP...01..047L}.

We have considered a flat cosmology with $\Omega_m = 0.31$, and $h = 0.68$ which are consistent with the 
recent Planck results \citeyearpar[Planck Collaboration XIII][]{2016A&A...594A..13P}. The absolute magnitude 
distribution of the SNe is taken from \citet{2014AJ....147..118R} and has been adjusted for the Hubble 
parameter value $h = 0.68$. We also put a conservative cut-off at 2.5$\sigma$ in the absolute magnitude 
distribution as there are no data available beyond 2.5$\sigma$ in the sample of \citet{2014AJ....147..118R}. 
We take the same IMF, $\dot\rho_\star(z)$ and the dust correction as given by \citet{2009JCAP...01..047L}. 
The progenitor mass range for the Type Ia SNe is taken to be 3\,--\,8 M$_{\odot}$ and for CCSNe it is taken 
to be 8\,--\,50 M$_{\odot}$ (mass range for all four CCSNe types i.e. Ibc, IIL, IIP and IIn). 
Further, for CCSNe the form $F(\lambda)$ is taken as the one listed in \citet{2009JCAP...01..047L}, 
and for Type Ia SNe we choose  $F(\lambda)$ as a 15000 K blackbody spectrum with cut-off 
due to UV blanketing at $\lambda < 4000 \ \mathring{A}$ \citep{1999A&A...350..349D,2011ApJ...729...55F}. 
For the calculation of magnitude limited detection rate we have used the expected magnitude limits 
($m_{lim,x}$) for the ILMT in the different spectral bands (see Section~\ref{Estimation1}).

In Figs.~\ref{fig_snr_cc} and \ref{fig_snr_1a}, the dotted-brown and the dashed-green lines show the 
cosmic supernova rate with and without dust correction, respectively. The solid-red lines correspond 
to the expected magnitude limited SN detection rate observed with the ILMT. The set of three 
magnitude limits respectively correspond to 1, 3 and 6 night integration time 
(see Section~\ref{Estimation1}).

The cut-off in the absolute magnitude distribution has a very strong effect on the high redshift 
cut-off seen in the magnitude limited detection rate plots. Changing the cut-off from 2.5$\sigma$ 
to 3.0$\sigma$ increases the cutoff redshift drastically, though it does not change the total 
supernovae counts by much and it has little effect over the redshift value corresponding to 
the maximum detection rate as has also been pointed out by \citet{2009JCAP...01..047L}. 
Setting a cut-off in the absolute magnitude distribution at 2.5$\sigma$ gives a good, 
yet conservative, estimate for the detection rate.

The sharp cut-off seen in the case of Type Ia SNe (Fig.~\ref{fig_snr_1a}, $g'$-band) is due to the UV 
blanketing effect that cuts-off the spectral energy distribution at $z \gtrsim \lambda/4000$. It makes the 
K-correction very high for $z \gtrsim \lambda/4000$ (see Fig.~\ref{fig_Kcor}; $g'$-band SN-Ia), making 
$\tilde{m}_{\rm \scriptscriptstyle Ia}(z \gtrsim \lambda/4000) \gg m_{lim,g'}$. 
As a result we get a sharp cut-off at a comparatively lower redshift in the $g'$-band for Type Ia SNe.


Ground based observing facilities are sometimes affected by local weather and the ILMT site is 
not an exception. The wet and humid conditions during the {\it Monsoon} season necessitate the
closure of all observing facilities situated at the site from June to September every year.
Therefore, ILMT observations will also be closed during that period (i.e. 4 months). Previous 
observing experiences suggest that in general, the Devasthal site has $\sim$210 clear nights in a year, 
among that $\sim$160 nights are of photometric quality \citep[see][]{2000A&AS..144..349S}. 
Taking into account the site photometric nights and the area covered by the ILMT each night,
we have estimated total number of possible SNe to be detected with this facility. 
These numbers are listed in Table~\ref{ILMT_sn_det} (see also Table~\ref{ILMT_sn_det_z}).
It is obvious that the above-estimated SNe numbers will reasonably vary during real observations if we 
also consider the night limitations and technical difficulties/maintenance of the instruments. 
The uncertainties in the above estimates are therefore not discussed further.

 
\begin{table}
\centering
\small
\caption{SN detection rates with the ILMT.
1$_{\rm N}$, 3$_{\rm N}$ and 6$_{\rm N}$ indicate the number of SNe for the limiting magnitudes of 
single, and co-added images of 3 and 6 nights, respectively.
Total number of SNe (columns 6, 7 and 8) are the redshift integrated
events in a year (only 160 photometric nights of the site and an average 8 hours of observing
time each night have been accounted for).} \label{ILMT_sn_det}
\begin{tabular}{cccccccc}
\hline
SN   &Filter   & \multicolumn{3}{c}{SNe/deg$^{2}$/year}  & \multicolumn{3}{c}{Total SNe in a year}\\
Type &   & 1$_{\rm N}$ & 3$_{\rm N}$ & 6$_{\rm N}$ & 1$_{\rm N}$ & 3$_{\rm N}$ & 6$_{\rm N}$ \\ \hline
Ia   & $g'$   &63 & 89 & 115 & 1299 & 1835 & 2371 \\
     & $r'$   &155 & 274 & 426 & 3196 & 5649 & 8783 \\
     & $i'$   &28 & 71 & 174 & 577 & 1464 & 3588 \\\\      
CC   & $g'$   &50 & 97 & 177 & 1031 & 2000 & 3649 \\
     & $r'$   &20 & 43 & 87 & 412 & 887 & 1794 \\
     & $i'$   &3 & 8 & 19 & 62 & 165 & 392 \\
\hline
\end{tabular}
\end{table}


\subsection{Detection of supernova candidates}

The ILMT pointing is fixed towards the best seeing and atmospheric transparency position (i.e. at zenith).
This allows one to obtain images with optimal quality. During each clear night, the same strip 
of sky will be scanned by the telescope. To detect SNe, previous night images or a good reference image 
will be subtracted from the search night images. We plan to perform automated real-time data reduction 
pipeline based on the Optimal Image Subtraction (OIS) technique \citep{1998ApJ...503..325A,
2000A&AS..144..363A}.

In order to detect SNe and classify their type, utmost care is needed. Miscellaneous astrophysical 
and/or non-astrophysical contaminating sources such as variable stars, quasars, active galaxies, 
asteroids, cosmic rays, etc. may appear on the acquired images. These unwanted sources must be removed 
accurately to avoid false detection. Various catalogues for different types of variable sources can be 
used to cross-match newly discovered transient candidates.
Some catalogues are: VERONCAT\footnote{\url{https://heasarc.gsfc.nasa.gov/w3browse/all/veroncat.html}}
\citep{2010A&A...518A..10V} for quasars and AGN, Minor Planet Checker\footnote{\url{http://www.minorplanetcenter.net}} 
for planets \& comets, and SIMBAD\footnote{\url{http://simbad.u-strasbg.fr/simbad/}} \citep{2000A&AS..143....9W}
for variable stars. Following the detection of a possible SN candidate, information will be communicated 
to the astronomical community (e.g. through ATEL and CBET), and also it will be available on the ILMT 
webpage.

The appearance of emission lines in the spectra is mandatory for classification and confirmation 
of a supernova. The spectrum is also useful to estimate the redshift and age (time after explosion) of 
SNe. However, distant SNe may be too faint for spectroscopy. Moreover, the follow-up spectroscopy 
will be a virtually impossible approach considering the large amount of transients detected in surveys. 
Therefore, new techniques have been developed attributing to SNe photometric classification.
These are primarily based on some form of light curve fitting models e.g.
MLCS/MLCS2k2\footnote{Multicolour light-curve shape} \citep{1995ApJ...438L..17R,2007ApJ...659..122J}
and SALT/SALT2\footnote{Spectral adaptive light curve template} \citep{2005A&A...443....1G,2007A&A...466...2G}.
The observed data points are fitted to the templates and the most likelihood type is determined
\citep[see, e.g.][]{2002PASP..114..833P,2007AJ....134.1285P,2004PASP..116..597G,2006AJ....131..960S,
2006AJ....132..756J,2007ApJ...659..530K,2007PhRvD..75j3508K,2009ApJ...707.1064R,2010ApJ...709.1420G,
2010ApJ...723..398F,2011ApJ...738..162S}. 
The above cited models along with colour cuts and colour evolution based techniques \citep[e.g.][]{2002PASP..114..284D,
2006AJ....132..756J} can be applied for the type determination of the ILMT discovered SNe.

A quick and dense monitoring (particularly, during the phase just after the explosion to a few weeks 
after the explosion) of SNe will help in the type identification more accurately. Furthermore, at the 
moment of shock break-out, some SNe are expected to emit a short burst of high energy radiation 
\citep{2010ApJ...725..904N}. Thereafter, the shock break-out cooling may create an early peak in the 
optical passband \citep{2013ApJ...769...67P}. The early phase observations are therefore crucial 
to constrain the progenitor system \citep{2012ApJ...757...31B,2015A&A...574A..60T}.  
Here we emphasize that the ILMT will work in the continuous data acquisition mode. 
Accordingly, while it will contribute with data on each night basis nonetheless,
it will not be possible to observe again a particular sky patch on the same night once it has passed
over the ILMT field of view. Additionally, due to the limitation of the ILMT filter system it will
be difficult to obtain precise colour and light curve information of SN candidates.
Therefore, complementary observations with conventional glass mirror telescopes will be very useful.
Thanks to the ARIES observing facilities which host three modern glass telescopes with different
apertures (the 1.04-m Sampurnanad Telescope ({\it ST}), the 1.3-m {\it DFOT} and 3.6-m {\it DOT}).
Depending upon the brightness and peculiarity of the newly discovered objects these telescopes along with 
other existing observing facilities in India and worldwide can be triggered for follow-up observations. 
A detailed follow-up scheme is presented in \citet{Kumar_bina}.

It is important to mention that along with the SN studies, the ILMT has other scientific interests also.
That includes surveys for multiply imaged quasars, determination of trigonometric parallaxes of faint 
nearby objects, detection of high stellar proper motions and short to long term photometric variability 
studies of stellar objects \citep[][]{2006SPIE.6267E...4S,Jean_bina}. Therefore, a reasonable 
balance between different filters is a must. The $i'$ filter can be placed around the bright moon 
phases and during the remaining nights a combination of $g'$, $r'$ and $i'$ filters can be placed
alternatively. Such an observing strategy may equally be useful for SN candidate detection and other 
science cases as well.


\section{Summary and conclusions}\label{concl}

The redshift-integrated supernova rate may turn out to be very large
\citep[$\simeq$5 -- 15 events/sec,][]{1998MNRAS.297L..17M}. Considering their random occurrence 
in the Universe, it is not feasible to detect and observe each event. Monitoring all of them is 
also almost impossible as it will require a significant amount of telescope time. 
On the other hand, a regular imaging of a same strip of sky with the ILMT will be 
advantageous to apply the image subtraction technique for detecting transients such as 
SNe. Moreover, once a SN is detected in the ILMT images, it will by default provide dense sampled 
(successive night) light curves in different filters.
The single scan ILMT limiting magnitudes are $\sim$22.8, $\sim$22.3 and $\sim$21.4 mag in $g'$, $r'$ 
and $i'$ filters, respectively and can be increased if we co-add the images taken on different nights.
In this way, the ILMT survey should play an important role in SNe detection up to reasonably 
fainter limits with a precise and unbiased imaging of a strip of sky at a declination equal 
to the latitude of Devasthal. 
We are expecting to detect hundreds of Type Ia as well as core-collapse SNe up to intermediate 
redshifts thanks to the ILMT survey (c.f. Table~\ref{ILMT_sn_det}).
The multi-band and well sampled observations should enable the photometric type determination 
(by template fitting, colour information) of SNe more accurately. The expected large SNe samples 
yielded from the ILMT may also increase the representative objects of each type with better 
statistics. Furthermore, the ILMT will provide an untargeted search with plentiful of anonymous 
galaxies in each night images, which may allow us to construct a SN sample without host-galaxy 
biases in a given limited patch of sky.

The observational properties and theoretical modelling indicate that supernova light curves are
mainly powered by a combination of two energy sources, i.e. shock generated energy deposition and
radioactive decay $^{56}$Ni $\rightarrow$ $^{56}$Co $\rightarrow$ $^{56}$Fe, synthesized in the
explosion \citep[see][]{1960ApJ...132..565H,1969ApJ...157..623C,1982ApJ...253..785A}. In some of the
SNe, circumstellar interactions \citep{1978MmSAI..49..389R,1991MNRAS.250..513C,1994ApJ...420..268C}
and rapidly rotating magnetars \citep{2010ApJ...717..245K,2010ApJ...719L.204W,2012MNRAS.426L..76D}
can also supply energy.
The light curves of different types of SNe exhibited diverse characteristics,
e.g. Type II SNe: \citep{2012ApJ...756L..30A,2014ApJ...786...67A,2015ApJ...799..208S},
stripped envelope SNe\footnote{In these events, the outer envelopes of hydrogen and/or helium 
of their progenitors are partially or completely removed before the explosion (e.g. Type IIb, Ib, 
Ic, and Ic-BL).}: \citep{2011ApJ...741...97D,2014ApJS..213...19B,2015A&A...574A..60T,
2016MNRAS.457..328L,2016MNRAS.458.2973P} and Type Ia SNe: \citep[][and references 
therein]{2010ApJ...712..350H,2016MNRAS.460.3529A,2017ApJ...846...58H}. 
The high quality light curves can provide an opportunity to robustly determine various parameters
and empirical relations including the `rise time', the light curve decline rate parameter
\citep[$\Delta$m$_{15}$,][]{1993ApJ...413L.105P}, and the color evolution etc. of different types 
of SNe. This will also allow us to use theoretical models for a robust determination of the explosion 
parameters (e.g. the $^{56}$Ni mass synthesized, the ejected mass and explosion energy). 
That will in turn shed some light on the explosion mechanisms of different SNe and evolutionary
stages of their progenitors.

It is noteworthy that the spectra of SNe provide crucial information about the composition and 
distribution of elements in the ejecta. Therefore, the spectroscopic monitoring of peculiar events 
will also be very valuable. In this context, a guaranteed-time allocation strategy on 3.6-m {\it DOT} 
to follow-up newly discovered objects will fulfill our needs.
Because of the tight link between SNe and star formation, the ILMT with complementary observations and 
along with other sky surveys (e.g. Large Synoptic Survey Telescope (LSST), ZTF, etc.) 
may provide better measurements of the moderate redshift history of the cosmic star-formation rate. 
New SNe discoveries and precise investigation of their light curves could improve our knowledge on a variety 
of problems including cosmology and SN physics.

\section*{Acknowledgments}
The authors thank the referee for his/her useful comments that substantially improved this paper.
We are grateful to the members of the ILMT team who provided their sincere efforts for this project
that is now in its final stage of installation. Although the list of ILMT contributors is long, we 
specially thank J. P. Swings, Ram Sagar, Hum Chand, Ludovic Delchambre, Serge Habraken and Bikram Pradhan
for their precious support in this project. 
Active involvement of Anna Pospieszalska is highly acknowledged.
BK and KLP also thank Amy Lien for fruitful discussions during the preparation of this manuscript.
BK acknowledges the Science and Engineering Research Board (SERB) under the Department of Science 
\& Technology, Govt. of India, for financial assistance in the form of National Post-Doctoral 
Fellowship (Ref. no. PDF/2016/001563).
SBP acknowledges BRICS grant DST/IMRCD/BRICS/Pilotcall/ProFCheap/2017(G) for the present work.
This research has also been supported by R\'{e}gion Wallonne (Belgium) under the convention 516238,
F.R.S.--FNRS (Belgium), the Li\`{e}ge University and the Natural Sciences and Engineering Research
Council of Canada. JS wishes to express his special thanks to Ir. Alain Gillin, Director,
for his comprehension in renewing the Convention no 516238 `TML4M' during many years and 
Prof. Govind Swarup, former Chair of the ARIES Governing Council, for his constant and very important
support. Part of this work was initiated during the doctoral thesis of Brajesh Kumar.

\bibliography{sn_rate}

\appendix 
\section{$K$-Correction}
\begin{figure*}
\centering
\includegraphics[width=\columnwidth]{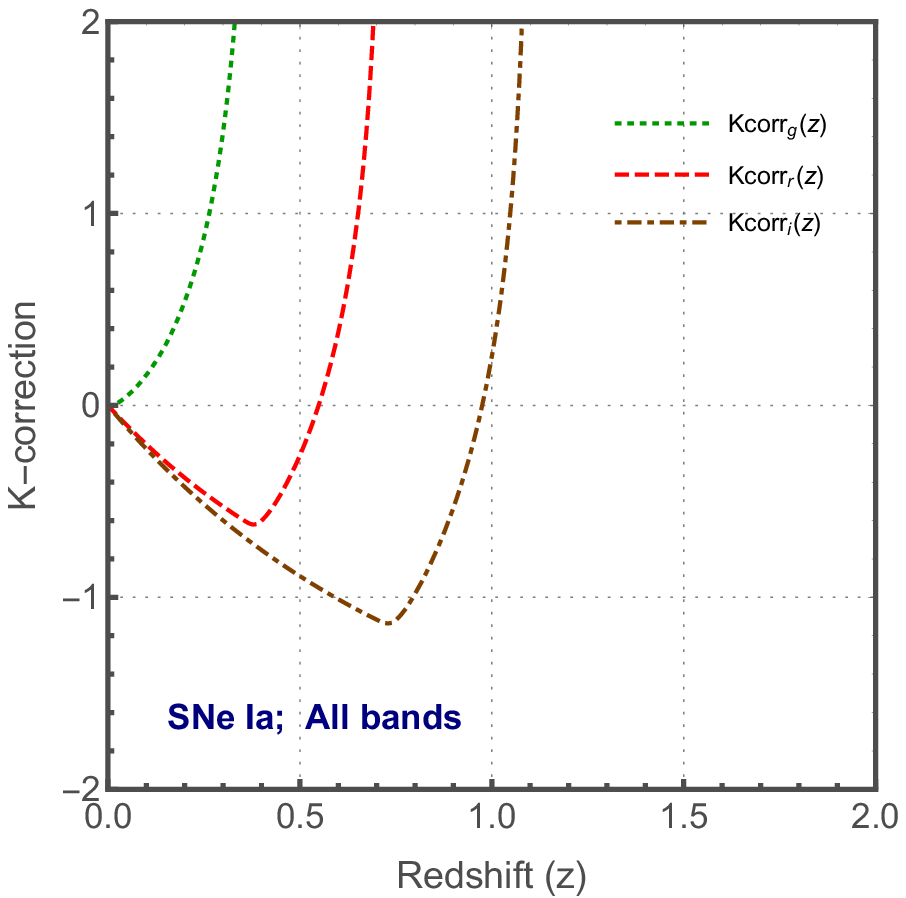}
\includegraphics[width=\columnwidth]{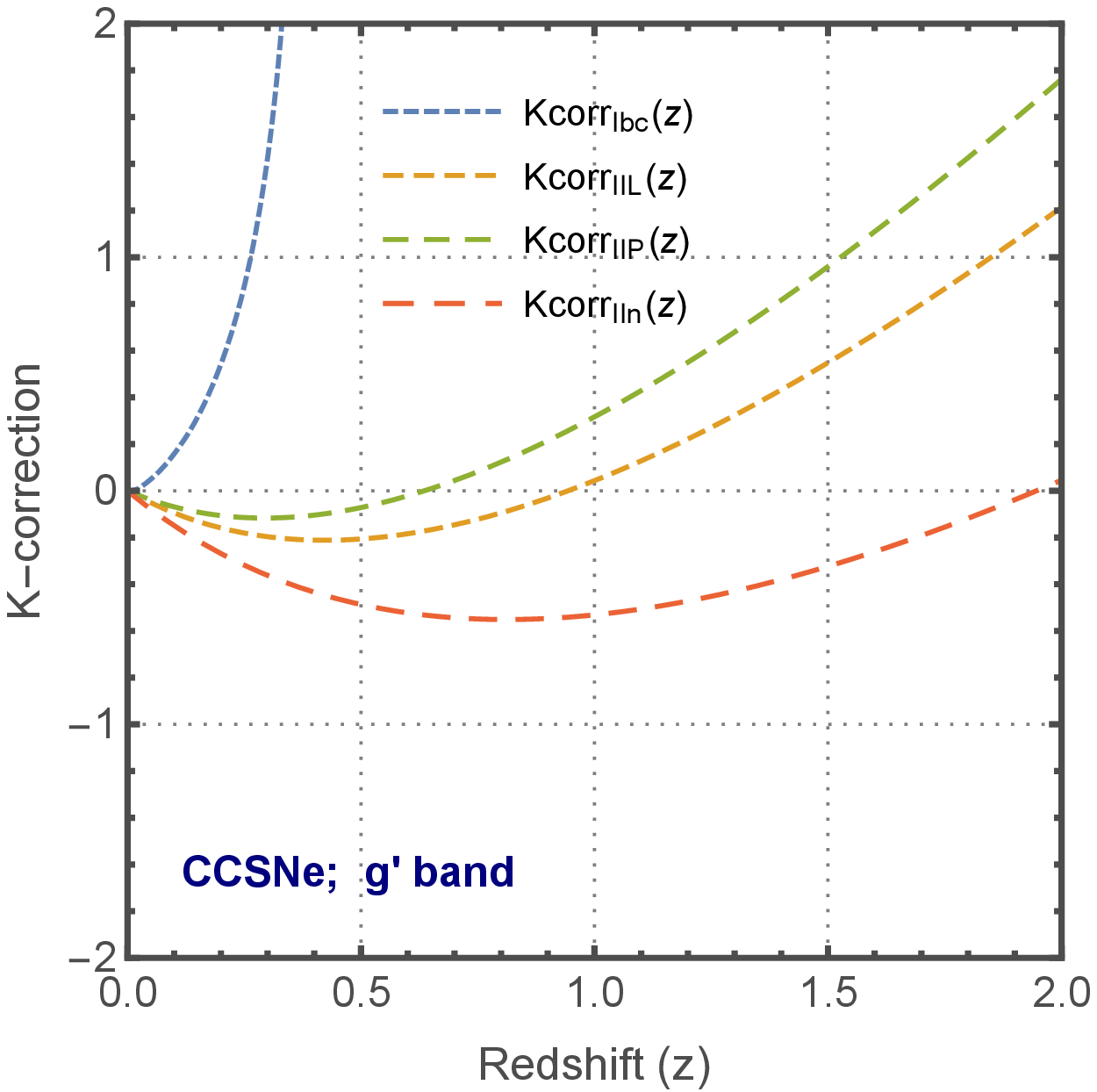}
\includegraphics[width=\columnwidth]{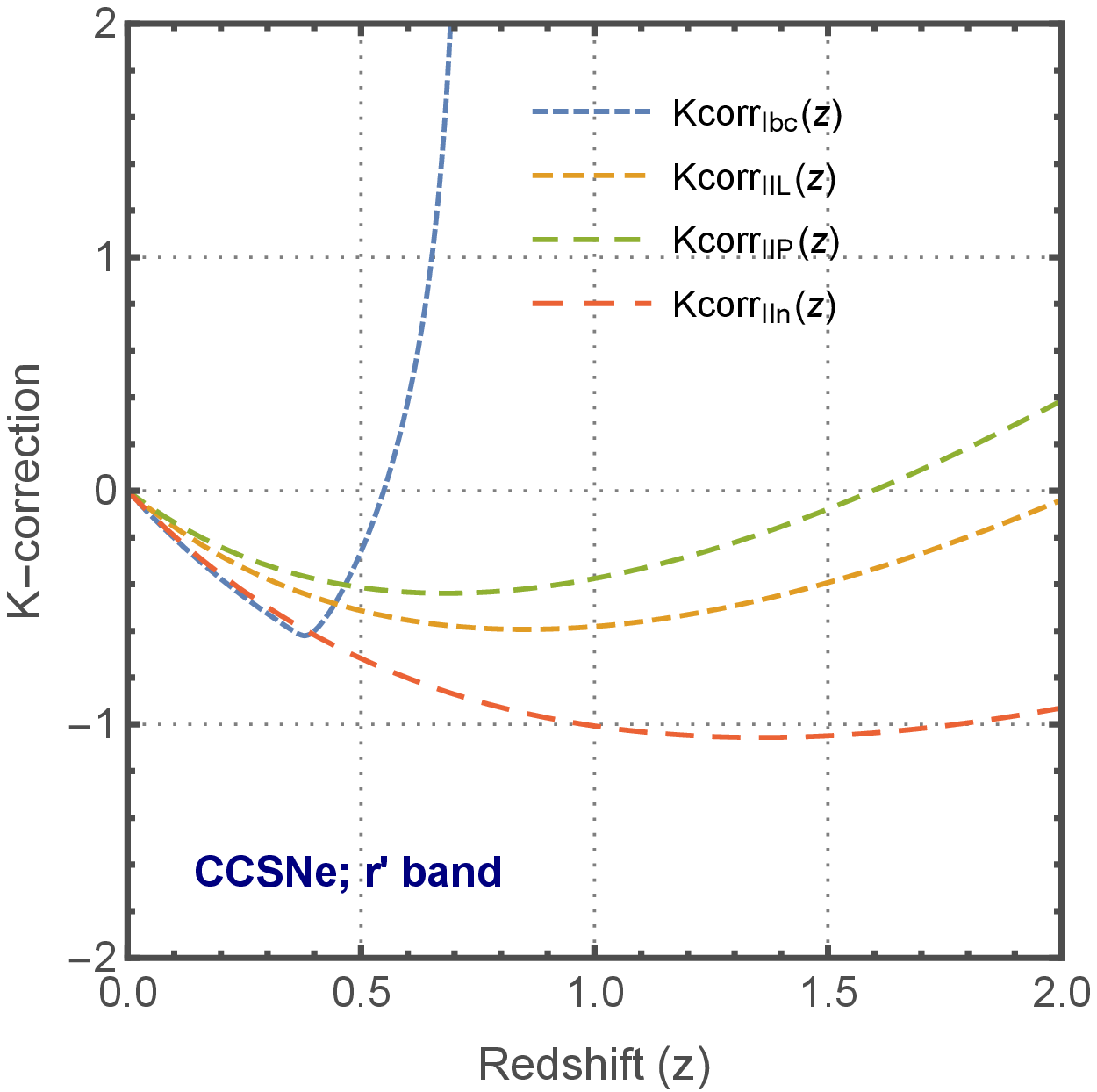}
\includegraphics[width=\columnwidth]{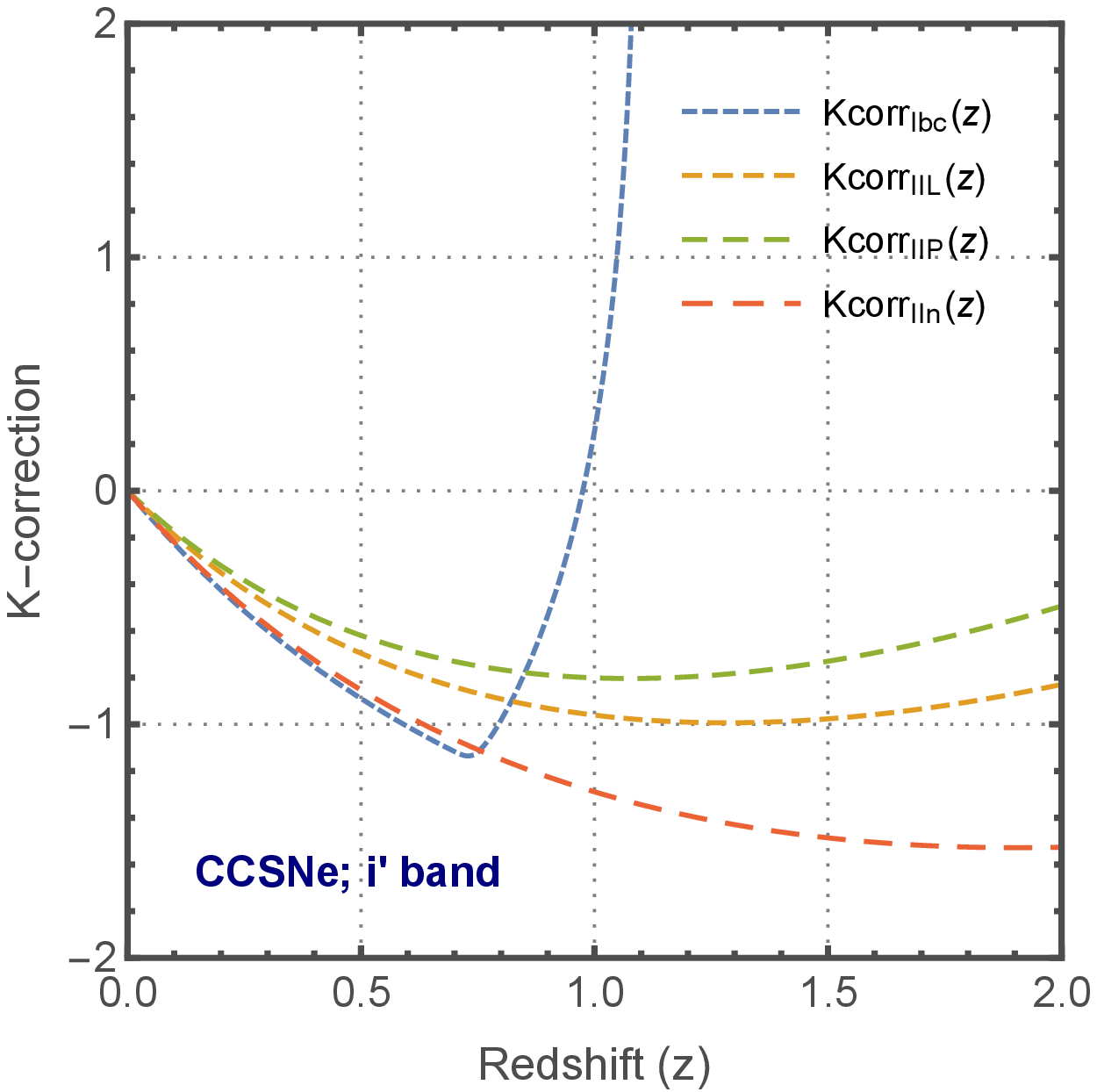}
\caption{$K$-correction plots for the different Types of SNe (Ia, Ibc, IIL, IIP and IIn) in $g'$, $r'$
and $i'$ bands.}
\label{fig_Kcor}
\end{figure*}

The K-correction term accounts for the conversion from an observed magnitude to that which would be observed
in the rest frame in another passband \citep{1956AJ.....61...97H,1968ApJ...154...21O}.
Fig.~\ref{fig_Kcor} represents K-corrections for different SNe that have been applied to calculate
ILMT SNe rate for the $g'$, $r'$ and $i'$ filters.

\section{Predicted SN with the ILMT}
In Table~\ref{ILMT_sn_det_z}, the predicted SN rate (for Ia and CC) estimated for $\Delta$$z$ = 0.2 bins are listed.

\begin{table}
\centering
\small
\caption{Predicted Type Ia and CCSNe (/deg$^{2}$/year) in $\Delta$$z$ = 0.2 bins.
1$_{\rm N}$, 3$_{\rm N}$ and 6$_{\rm N}$ indicate the number of SNe for the limiting magnitudes of
single, and co-added images of 3 and 6 nights, respectively.}
\label{ILMT_sn_det_z}
\begin{tabular}{cccccccc}
\hline\
Redshift   &Filter   & \multicolumn{3}{c}{Ia SNe}  & \multicolumn{3}{c}{CCSNe}\\
($z$)      &         & 1$_{\rm N}$ & 3$_{\rm N}$ & 6$_{\rm N}$ & 1$_{\rm N}$ & 3$_{\rm N}$ & 6$_{\rm N}$ \\ \hline
0.2   &        & 29 & 30   & 30    & 8  & 10 & 10   \\ 
0.4   &        & 63 & 89   & 115   & 31 & 44 & 57   \\
0.6   &        & 63 & 89   & 115   & 45 & 75 & 111  \\
0.8   & $g'$   & 63 & 89   & 115   & 50 & 91 & 150  \\
1.0   &        & 63 & 89   & 115   & 50 & 96 & 169  \\
1.2   &        & 63 & 89   & 115   & 50 & 97 & 176  \\
1.4   &        & 63 & 89   & 115   & 50 & 97 & 177  \\
\hline
0.2   &        & 29  & 30  & 30    & 06 & 08 & 09   \\  
0.4   &        & 133 & 199 & 243   & 16 & 29 & 44   \\
0.6   &        & 155 & 274 & 424   & 20 & 39 & 71   \\
0.8   & $r'$   & 155 & 274 & 426   & 20 & 42 & 82   \\
1.0   &        & 155 & 274 & 426   & 20 & 43 & 86   \\
1.2   &        & 155 & 274 & 426   & 20 & 43 & 87   \\
1.4   &        & 155 & 274 & 426   & 20 & 43 & 87   \\
\hline
0.2   &        & 17  & 24  & 28    & 02 & 04 & 05   \\ 
0.4   &        & 28  & 63  & 122   & 03 & 07 & 14   \\
0.6   &        & 28  & 71  & 165   & 03 & 08 & 18   \\
0.8   & $i'$   & 28  & 71  & 174   & 03 & 08 & 19   \\
1.0   &        & 28  & 71  & 174   & 03 & 08 & 19   \\
1.2   &        & 28  & 71  & 174   & 03 & 08 & 19   \\
1.4   &        & 28  & 71  & 174   & 03 & 08 & 19   \\
\hline
\end{tabular}
\end{table}

\label{lastpage}
\end{document}